\documentclass[usegraphicx,usenatbib]{mn2e}
\usepackage{amsmath}
\usepackage{amssymb}
\usepackage{mathptmx}
\usepackage{subfigure}
\usepackage{epsfig}
\usepackage{bm}

\renewcommand{\descriptionlabel}[1]%
  {\hspace{\labelsep}\textbf{#1}}

\title[Investigation of variable star candidates in NGC~5024]
      {Investigation of variable star candidates in the globular cluster NGC~5024 (M53)}

\author[D.M. Bramich et al.]
  {D. M. Bramich$^{1}$\thanks{E-mail: dbramich@eso.org, dan.bramich@hotmail.co.uk},
   A. Arellano Ferro$^{2}$,
   R. Figuera Jaimes$^{2}$
   and Sunetra Giridhar$^{3}$
  \medskip
  \\$^{1}$European Southern Observatory, Karl-Schwarzschild-Stra$\beta$e 2, 85748 Garching bei M\"{u}nchen, Germany
  \\$^{2}$Instituto de Astronom\'ia, Universidad Nacional Aut\'onoma de M\'exico, M\'exico
  \\$^{3}$Indian Institute of Astrophysics, Koramangala 560034, Bangalore, India
  }

\begin{document}

\date{Accepted 2010 August ???. Received 2010 August ???; Submitted 2010 August ???}

\pagerange{\pageref{firstpage}--\pageref{lastpage}} \pubyear{2010}

\maketitle

\label{firstpage}

\begin{abstract} 
We have performed a careful investigation of the 74 candidate variable stars presented by \citet{saf2011}.
For this purpose we used our data base of imaging and light curves from \citet{are2011} and \citet{are2012}. We find that two
candidates are known variable stars, eight candidates were discovered first by \citet{are2011} but would not have been
known to \citet{saf2011} at the time of their paper submission, while four candidates are new variables.
Three of the new variables are SX Phe type and
one is a semi-regular red giant variable (SR type). We also tentatively confirm the presence of true variability
in two other candidates and we are unable to investigate another four candidates because they are not in our data base.
However, we find that the remaining 54 candidate variable stars are spurious detections where systematic trends in the
light curves have been mistaken for true variability. We believe that the erroneous detections are caused by the
adoption of a very low detection threshold used to identify these candidates.
\end{abstract} 


\section{Introduction}
\label{sec:introduction}

Performing a census of variable stars in a globular cluster is an important task because variable stars, such as RR Lyrae and SX Phe, provide
a wealth of information about the host cluster. Fourier decomposition of the RR Lyrae light curves can be used to estimate their metallicity and
absolute magnitudes, from which estimates of the metallicity and distance to the cluster may be derived. A frequency analysis of the light curves of the
SX Phe stars combined with the knowledge of their period-luminosity (P-L) relation may also be used to determine a cluster distance, and conversely,
the results may be used to investigate how the SX Phe P-L relation depends on metallicity (\citealt{coh2012}).

When performing precise differential time-series photometry, essential for a variable star census, there is always the problem of systematic photometric
errors that correlate with image and object properties such as point-spread function full-width half-maximum (PSF FWHM), spatial coordinates,
local sky background, etc. (``red noise'' - see \citealt{pon2006}; \citealt{bra2012}).
Systematic trends may sometimes be mistaken for real variability, since image and object properties may vary smoothly over time, mimicking variability on 
various timescales. Therefore it can be quite easy to contaminate the catalogue of true variable stars in a globular cluster with non-variable stars
whose light curves exhibit systematic trends, and consequently these stars may be accidentally used as part of the sample of variable stars to derive the cluster 
properties, introducing unwanted errors into these estimates. One way to avoid contamination is to set the detection threshold for variables to a level that
has a very small probability of admitting false detections even when systematic trends are present. The detection threshold may be determined empirically from an analysis of 
the light curves of the general ensemble of stars which are usually dominated by true non-variable stars.

In a series of papers (e.g. \citealt{are2008}; \citealt{are2010}; \citealt{bra2011}) we have been performing a variable star census for a range of Galactic globular clusters
of different metallicities. We have targeted clusters that are generally under-studied with regards to time-series photometry in the CCD era, and we aim to detect all the variables in the
cluster field-of-view down to close to our magnitude limit. To avoid confusion over the variable star identifications for future researchers, we provide detailed
finding charts and accurate astrometry in addition to the light curves of the variable stars that we find.

Specifically, we recently performed this task for NGC~5024 (\citealt{are2011} - from now on A11; \citealt{are2012} - from now on A12).
We used the 2.0~m telescope of the Indian Astronomical Observatory (IAO), Hanle, India,
equipped with an imaging camera with a pixel scale of 0.296 arcsec pix$^{-1}$ and a field-of-view (FOV) of $\sim$10.1$\times$10.1 arcmin$^{2}$, to obtain $V$ and $I$ time-series photometry with $\sim$300
epochs in each waveband observed during 18 nights spread out over $\sim$2 years. Our exposure times ranged from 25~s to 600~s in both filters, with the
longer exposure times generally employed during the earlier observation nights. We reduced the image data using the 
technique of difference image analysis (DIA) and specifically we used
the {\tt DanDIA} software (\citealt{bra2008}) for this purpose. We clarified the variable status, periods, and ephemerides of the
previously known variables, including correcting the misidentification of three variables, and we presented the discovery of two new RR Lyrae stars and 13 new SX Phe stars.

More recently, \citet{saf2011} (from now on SS11) used the same telescope and camera to obtain $\sim$7.2 hours of continuous $R$-filter time-series photometry (101 epochs) 
from a single night and using a single exposure time of 100~s. We therefore expect that the SS11 image properties are very similar to those of the A11 and A12 images (e.g. 
cluster positioning, seeing, number of saturated stars, etc.). SS11 also reduced their image data using {\tt DanDIA}, although the parameters used for the reduction process were not necessarily 
the same as those used in A11 and A12. SS11 claim the detection of 74 new variable stars from the analysis of their data. The purpose
of this paper is to investigate these new candidate variable stars in order to confirm their variable status using our more extensive photometric data base.
We also take this opportunity to publish some corrections of mistakes in the literature regarding variable star identification and astrometric coordinates in Appendix~A.

\section{Investigation Of The SS11 Variable Star Candidates In NGC~5024}
\label{sec:clar}

A total of 74 stars are presented as candidate variables in SS11; namely, one standard star exhibiting suspected eclipses, 14 RR Lyrae-type variables, 
10 eclipsing binaries, 22 SX Phe-type variables, and 27 other unclassified variables. In this Section, we investigate each set of candidate variables in turn.

\subsection{Variability In Standard Stars}
\label{sec:std_stars}

SS11 note that six Stetson standard stars (\citealt{ste2000}) exhibit spurious variability with the same pattern in their light curves, which they explain as
being caused by a deficiency in the data reduction process. However, star S240 is also suspected of showing the signature of
an Algol-type detached eclipsing binary with two suspected eclipses of depths $\sim$0.05 and $\sim$0.1~mag separated in time by $\sim$2.5~h, although
ultimately SS11 classify the variability seen in their light curve as spurious (see their Table~5).

We plot our $V$ light curve\footnote{The $I$ light curve does not exist in our data for this star because it lies on a bad column in the $I$ filter
reference image.} for S240 in Figure~\ref{fig:nonvar1} using the same scale on the magnitude axis as that used in SS11. Our light curve shows some systematic
photometric trends at the level of $\sim$0.01-0.02~mag and we achieve a root-mean-square (RMS) magnitude deviation in the light curve of $\sim$9~mmag. Given that our light curve
includes multiple sections of continuous time-series photometry of durations greater than 4~h, we should be able to detect an eclipsing signal with period $\sim$2.5~h
and amplitude $\sim$10 times our light curve noise. However, we detect no such signal, and we conclude that the eclipse-like signatures seen in the SS11 light curve
of star S240 are artefacts of the data reduction process.

\subsection{RR Lyrae-Type Variable Candidates}
\label{sec:rrlyrae}

\begin{figure*}
\centering
\epsfig{file=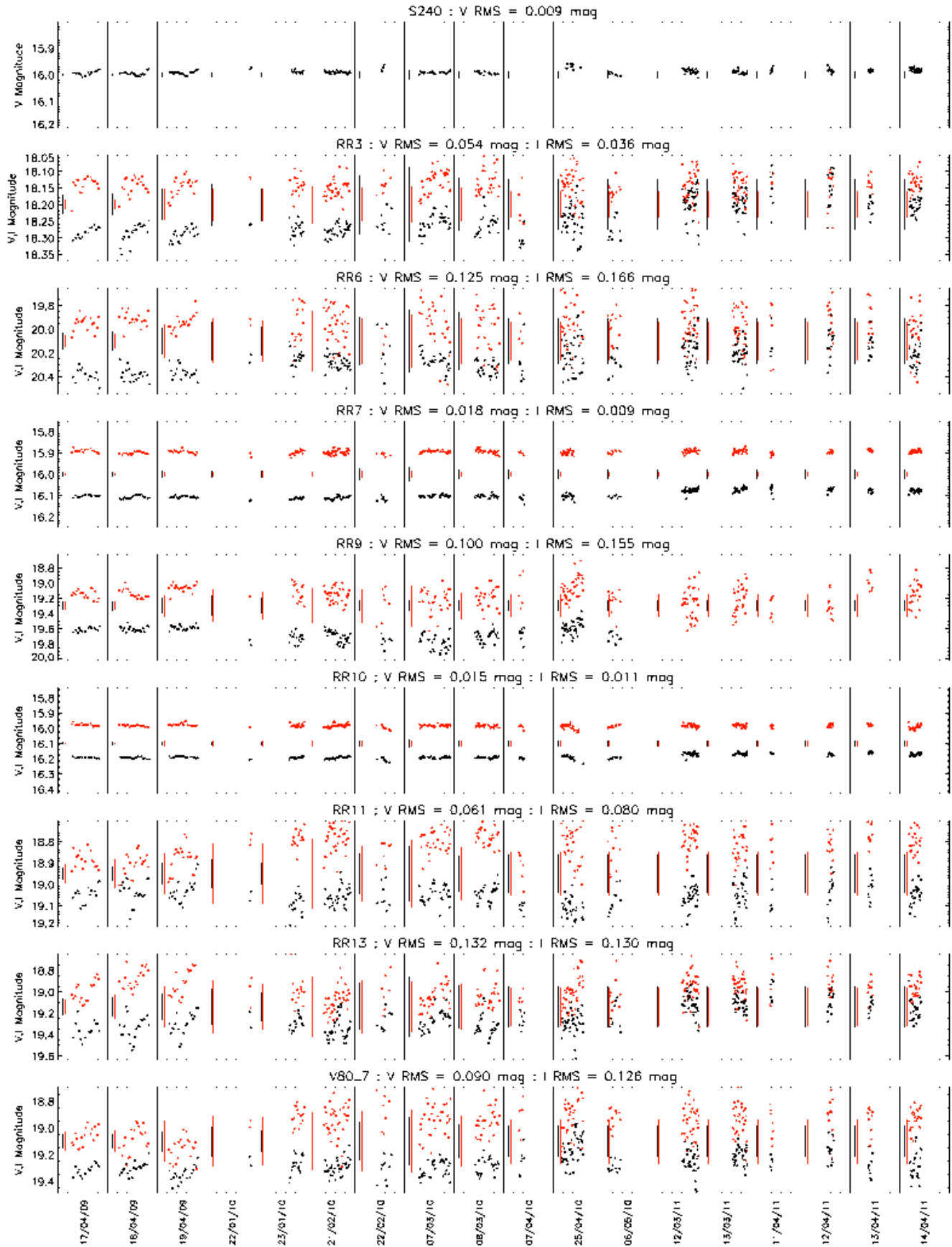,angle=0.0,width=\linewidth}
\caption{Light curves from A11 and A12 for selected candidate variables from SS11. $V$ and $I$ filter light curves are plotted in black and red, respectively,
         with mean photometric uncertainties per data point on each night plotted at the beginning of a night as vertical bars. The magnitude
         scale for each candidate has been matched to the scale used in SS11, and the $I$ filter light curves have been offset in magnitude by an arbitrary amount  
         for clarity.
         \label{fig:nonvar1}}
\end{figure*}

As described in \citet{bra2011}, the {\tt DanDIA} software measures
the differential flux of each star in the difference images by optimally scaling the appropriate point-spread function (PSF) model
at the known star position (as determined from the reference image) while simultaneously fitting for a local background.
The software stores the PSF model as a circular image stamp of radius four times the FWHM and it follows that for star pairs in close 
spatial proximity their corresponding PSF models may overlap. However, the {\tt DanDIA} software only performs an independent fit to each star in a difference image
and consequently the differential flux from a real variable may systematically influence the results of the PSF fit to any nearby stars,
with the effect being largest for the closest and faintest nearby stars. An easy way to spot if the flux variations from a nearby variable
star have influenced the photometry of an otherwise constant star is to see if the light curves of both stars correlate, or if they have the same period and phase.

SS11 present 14 candidate variables that are suspected of showing RR Lyrae-type light curve variations. However, we find that candidates RR1, RR2, RR8, RR12, and RR14 lie within
$\sim$7.6, 3.8, 16.9, 11.4, and 20.2~pix, respectively, of the known RR Lyrae variables V11, V18, V40, V1, and V58, respectively. Given that the typical PSF FWHM
of the SS11 observations is $\sim$5~pix ($\sim$1.5~arcsec), these candidate variables have PSFs that overlap substantially with the nearby RR Lyrae star PSFs. Furthermore,
these candidate variables are fainter than the known RR Lyrae variables (by $\sim$1.5-2.5~mag) and therefore we believe that the RR Lyrae-like variations that
are seen in the SS11 light curves of RR1, RR2, RR8, RR12, and RR14 are due to the systematic influence on the light curves from the nearby RR Lyrae stars.

Using the celestial coordinates from SS11 of the remaining RR Lyrae-type variable candidates, we have identified these stars in our $V$ and $I$ reference images and
found the corresponding light curves in our data base. Candidates RR4 and RR5 are very faint in our reference images and we do not have the corresponding light curves.
We cannot therefore comment on the variations seen in these stars by SS11.
For RR3, RR6, RR7, RR9, RR10, RR11, and RR13, we plot our light curves in Figure~\ref{fig:nonvar1} using black and red points for the $V$ and $I$ photometric measurements,
respectively, and using the same scale on the magnitude axis for each candidate variable as that used in SS11 to enable a direct comparison.
The $I$ light curves have been offset in magnitude by an amount that allows a clear comparison of the light curves between the two filters.
Note that the scatter in our light curves increases with time simply because our adopted exposure times were systematically reduced throughout our
observing campaign.

True photometric variability generally correlates between different filters whereas systematic trends in the photometry do not necessarily correlate but they may do. Furthermore,
systematic trends tend to exhibit similar temporal patterns between stars that are relatively close to each other in an image. Candidates RR11 and RR13 are within 50 pix of 
each other and their $V$ light curves show similar trends with time, which is also the case for the $R$ lightcurves of these stars displayed in Figure~11 of SS11. Also, 
these candidate variables are very close to a star that is highly saturated in both of our reference images which has a negative impact on the quality of the difference 
images in this region, explaining the particularly large amplitude of the trends in our light curves for these stars. Later on in Section~\ref{sec:unclass}, we also conclude that the
variable candidates VC18, VC19, and VC20 exhibit spurious variability due to their proximity to the same highly saturated star. For these reasons we do not believe that
SS11 have detected true variability in stars RR11 and RR13.

Our light curves for candidate variables RR7 and RR10 rule out the existence of the $\sim$0.2-0.3~mag amplitude variations seen in the SS11 light curves of these stars.
Inspection of our light curves for RR3, RR6, and RR9 reveals that the trends present in our light curves do not generally correlate between filters (e.g. see the night of 17/04/2009),
and certainly the trends do not reach the $\sim$0.5 and 0.4~mag amplitudes seen in the SS11 light curves for RR6 and RR9, respectively. We therefore believe that the larger-amplitude
variations seen by SS11 and the smaller-amplitude variations seen in our light curves for these five candidates are due to trends introduced during the reduction process.

Finally, in this section we comment that all of the RR Lyrae-type variable candidates RR1-RR14 as reported by SS11 are fainter ($R \sim 18-20$~mag) than the horizontal
branch of NGC~5024 ($R \sim 17.2$~mag), which implies that if they are RR Lyrae stars, then they must lie behind the cluster relative to our Sun, and therefore they cannot
be cluster members. It is highly unlikely that 14 Galactic RR Lyrae stars not associated with NGC~5024 lie within this $\sim$10$\times$10~arcmin$^{2}$ field of 
view. This conclusion is supported by the fact that over an area of $\sim$10000~deg$^{2}$, \citet{ses2011} found on average only $\sim$0.5 RR Lyrae stars per
square degree down to $R \sim 18$~mag. If any of these stars are eventually found to be true variables, then they are most likely to be of the SX Phe type.

\subsection{Eclipsing Binary Variable Candidates}
\label{sec:eclbin}

\begin{figure*}
\centering
\epsfig{file=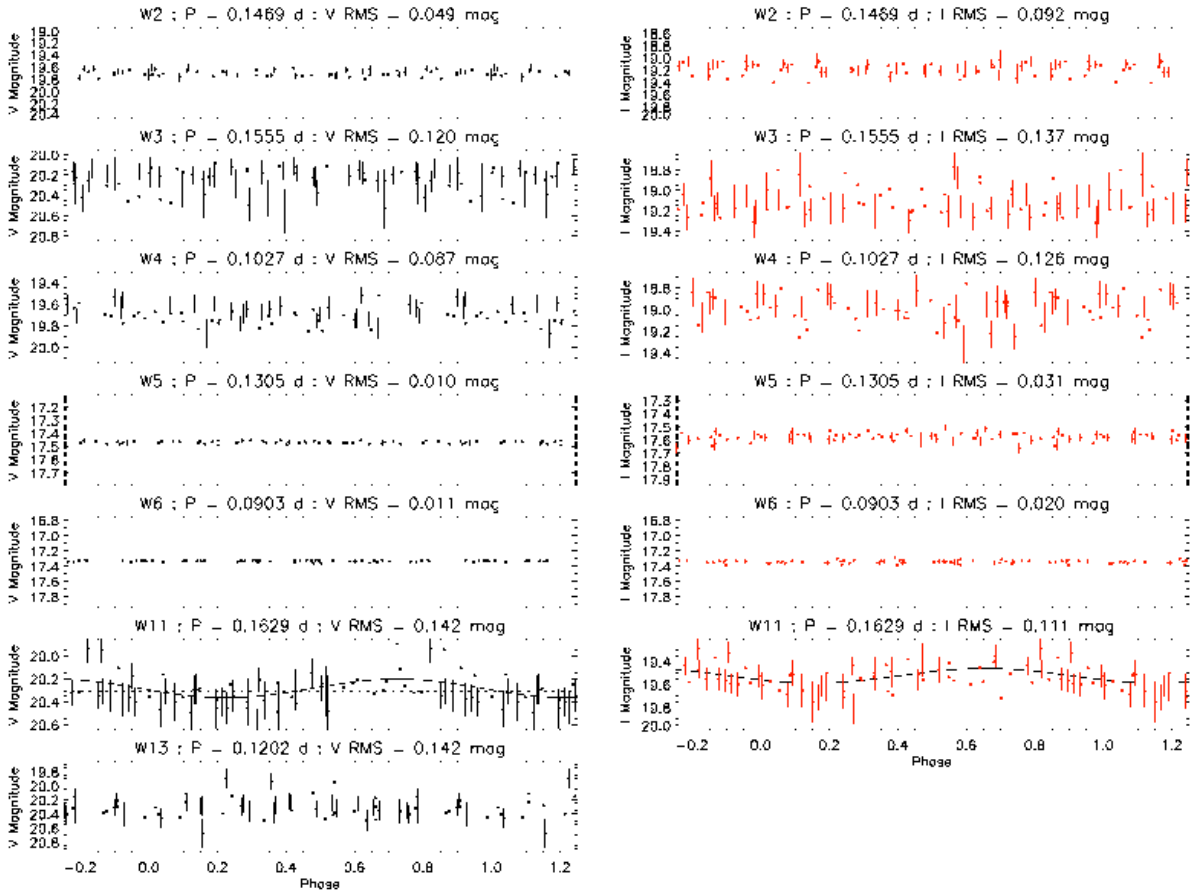,angle=0.0,width=0.9\linewidth}
\caption{Light curves of the first three nights of observations from A11 and A12 for selected candidate variables from SS11. The light curves have been phase-folded using the 
         periods derived by SS11. $V$ and $I$ filter light curves are plotted in the left and right-hand panels, respectively. The magnitude
         scale for each candidate has been matched to the scale used in SS11. For the candidate W11 showing true variability, fitted constant and
         sine-curve models are plotted as dashed and continuous curves, respectively.
         \label{fig:eclbin}}
\end{figure*}

SS11 present 10 candidate variables that they classify as short-period contact eclipsing binaries. They determine a period for each candidate and present the
phase-folded light curves. Using the celestial coordinates from SS11, we have found the corresponding $V$ and $I$ light curves in our data base.

Candidates W1 and W8 lie within $\sim$14.2 and 15.4~pix of the known variables V33 (RR Lyrae type) and V68 (SR type - red giant; highly saturated in our data), and for 
the same reasons as described in Section~\ref{sec:rrlyrae}, we do not believe that SS11 have detected true variability in these stars.
However, candidate W9 is a true variable star, but it is not an eclipsing binary. Instead it is the SX Phe type variable V104 first detected by A11.
SS11 derive a tentative period of $\sim$0.15~d for V104 consistent with the much more precise period of $\sim$0.14756~d from A11.

The periods determined by SS11 for any of their variable candidates are not appropriate for folding our light curves over the full time span of our observations ($\sim$2 
years) since they are derived from a time base of only $\sim$7.2 hours. Therefore, whenever we choose to plot phase-folded versions of our light curves using the
periods from SS11, we only plot the photometric data from the first three nights of observation (17-19th April 2009) since these nights are consecutive and the data have the best
photometric precision in our time-series because they correspond to the longest exposure times of 600~s in $V$ and 420~s in $I$.

In Figure~\ref{fig:eclbin}, we plot our light curves for candidates W2-W6, W11, and W13, phase-folded using the periods from SS11. We do not have an $I$ light curve
for candidate W13 because this star lies on a bad column in the $I$ filter reference image. SS11 list the amplitudes of the light curve variations in the $R$ band as
$\sim$1.22, 0.49, 0.75, 0.30, 0.60, 0.54, and 1.06~mag for the candidates W2, W3, W4, W5, W6, W11, and W13, respectively, and these variations, if true, will have
similar amplitudes in the $V$ and $I$ bands. Inspection of our plotted light curves reveals that we can fully rule out the presence of these variations in the candidates 
W2, W4, W5, W6, and W13. For W3, our light curves show a relatively large scatter, but it is clear that there are no coherent variations at the SS11 period.
We therefore believe that the variations seen by SS11 are due to trends introduced during the reduction process.

For the variable candidate W11, our phase-folded light curve in Figure~\ref{fig:eclbin} shows a hint of coherent variations at the SS11 period that match the shape of the
SS11 phase-folded light curve, and therefore we tentatively confirm the variable nature of this star. However, the scatter in our light curve is sufficiently large that
we believe that better precision follow-up observations are required before this candidate variable is assigned a proper V-number. We note that W11 has $V \approx 20.306$~mag and
$V - I \approx 0.763$~mag, which places it outside the blue straggler region as marked in the colour-magnitude diagram (CMD; Figure~4 in A11).
We also find that its period and mean magnitude are inconsistent with the P-L relation for SX Phe cluster members (see later in Figure~\ref{fig:plrel}),
since the star is much too faint for such a long period\footnote{W11 is not plotted in Figure~\ref{fig:plrel}.}.
We therefore cannot speculate on the type of variability exhibited by W11.

To quantify our conclusions about the candidate variables plotted in Figure~\ref{fig:eclbin}, we calculate the improvement in chi-squared $\Delta \chi^{2}$
when fitting a sine-curve compared to a constant magnitude. Under the null hypothesis that a light curve is not variable, the $\Delta \chi^{2}$
statistic follows a chi-squared distribution with two degrees of freedom. We set our threshold for rejection of the null hypothesis at 1\%, which is
equivalent to $\Delta \chi^{2} \ga 9.21$. Candidate W11 has $\Delta \chi^{2} \approx$~9.33 and 9.39 for the $V$ and $I$ light curves, respectively,
supporting our tentative conclusion that it is variable. We plot the fitted constant and sine-curve models in the corresponding panel of
Figure~\ref{fig:eclbin} as the dashed and continuous curves, respectively. For the remaining candidates shown in Figure~\ref{fig:eclbin},
the maximum value of $\Delta \chi^{2}$ is $\sim$4.42, meaning that we do not detect any sinusoidal variability in our data at the SS11 periods.

\subsection{SX Phe-Type Variable Candidates}
\label{sec:sxphe}

\begin{figure*}
\centering
\epsfig{file=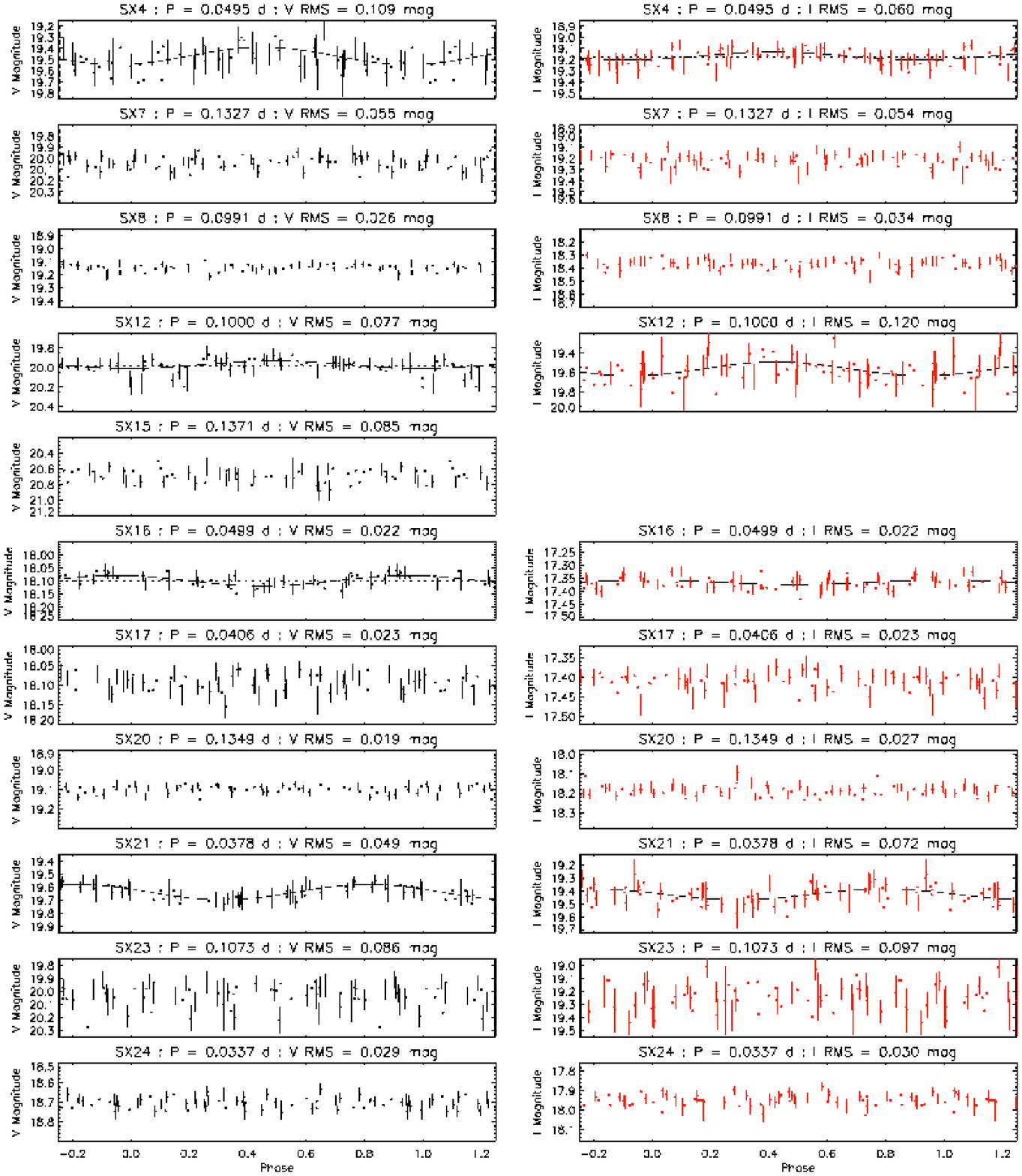,angle=0.0,width=0.9\linewidth}
\caption{Light curves of the first three nights of observations from A11 and A12 for selected candidate variables from SS11. The light curves have been phase-folded using the
         periods derived by SS11. $V$ and $I$ filter light curves are plotted in the left and right-hand panels, respectively. The magnitude
         scale for each candidate has been matched to the scale used in SS11. For the candidates showing true variability, fitted constant and
         sine-curve models are plotted as dashed and continuous curves, respectively.
         \label{fig:sxphe}}
\end{figure*}

SS11 present 22 candidate variables that they classify as ``potential SX Phe stars''. They determine a period for each candidate and present the
phase-folded light curves. Again, using the celestial coordinates from SS11, we have found the
corresponding $V$ and $I$ light curves in our data base, except for the candidate SX13 where we cannot find a star at the SS11 coordinates in
our $V$ and $I$ reference images.

Candidate SX6 is the known SX Phe variable star V78 first discovered by \citet{dek2009}. Candidates SX2, SX3, SX9\_1, SX11, SX14, SX19, and SX22 are
the SX Phe variable stars V96, V97, V99, V101, V102, V103, and V105 first discovered by A11. Candidate SX9, as noted by SS11, is only
$\sim$4.2~pix from the true variable V99, and its SS11 photometry has obviously been influenced by the flux variations from V99
because SS11 derive very similar periods and the same epoch for both stars. Also, as acknowledged by SS11, candidate SX25 is the
known SX Phe variable star V80 first discovered by \citet{dek2009}.

In Figure~\ref{fig:sxphe}, we plot our light curves for the candidates SX4, SX7, SX8, SX12, SX15-17, SX20, SX21, SX23, and SX24, phase-folded using the
periods from SS11. We do not have an $I$ light curve for the candidate SX15 due to a failure of the {\tt DanDIA} software to measure the
reference flux of this star in the $I$ filter reference image. Applying the same chi-squared analysis to these folded light curves as that
described in Section~\ref{sec:eclbin}, we find that SX4 has $\Delta \chi^{2} \approx$~14.43 and 10.22, SX12 has
$\Delta \chi^{2} \approx$~8.18 and 10.86, SX16 has $\Delta \chi^{2} \approx$~23.05 and 3.79, and SX21 has $\Delta \chi^{2} \approx$~34.67 and 11.59,
all for the $V$ and $I$ filters, respectively. We therefore confirm the variability at the SS11 periods for the candidates SX4, SX12, SX16, and SX21,
although we consider the candidate SX12 as border-line because the two $\Delta \chi^{2}$ values for the light curve lie only just above and below the
detection threshold. We plot the fitted constant and sine-curve models in the corresponding panels of Figure~\ref{fig:sxphe} as the dashed and continuous curves, respectively.

For the remaining candidates SX7, SX8, SX15, SX17, SX20, SX23, and SX24, the maximum value of $\Delta \chi^{2}$ is $\sim$3.51, and we conclude that
our data do not show any variability at the SS11 periods. Also, if the variations detected by SS11 in their light curves
for these stars are due to true variability, then we should have been able to detect them since their measured amplitudes are considerably larger than the scatter
in our light curves in all cases. Therefore we believe that the variations seen in the SS11 light curves of these candidates are due to trends
introduced during the reduction process.

\subsection{Unclassified Variable Candidates}
\label{sec:unclass}

\begin{figure*}
\centering
\epsfig{file=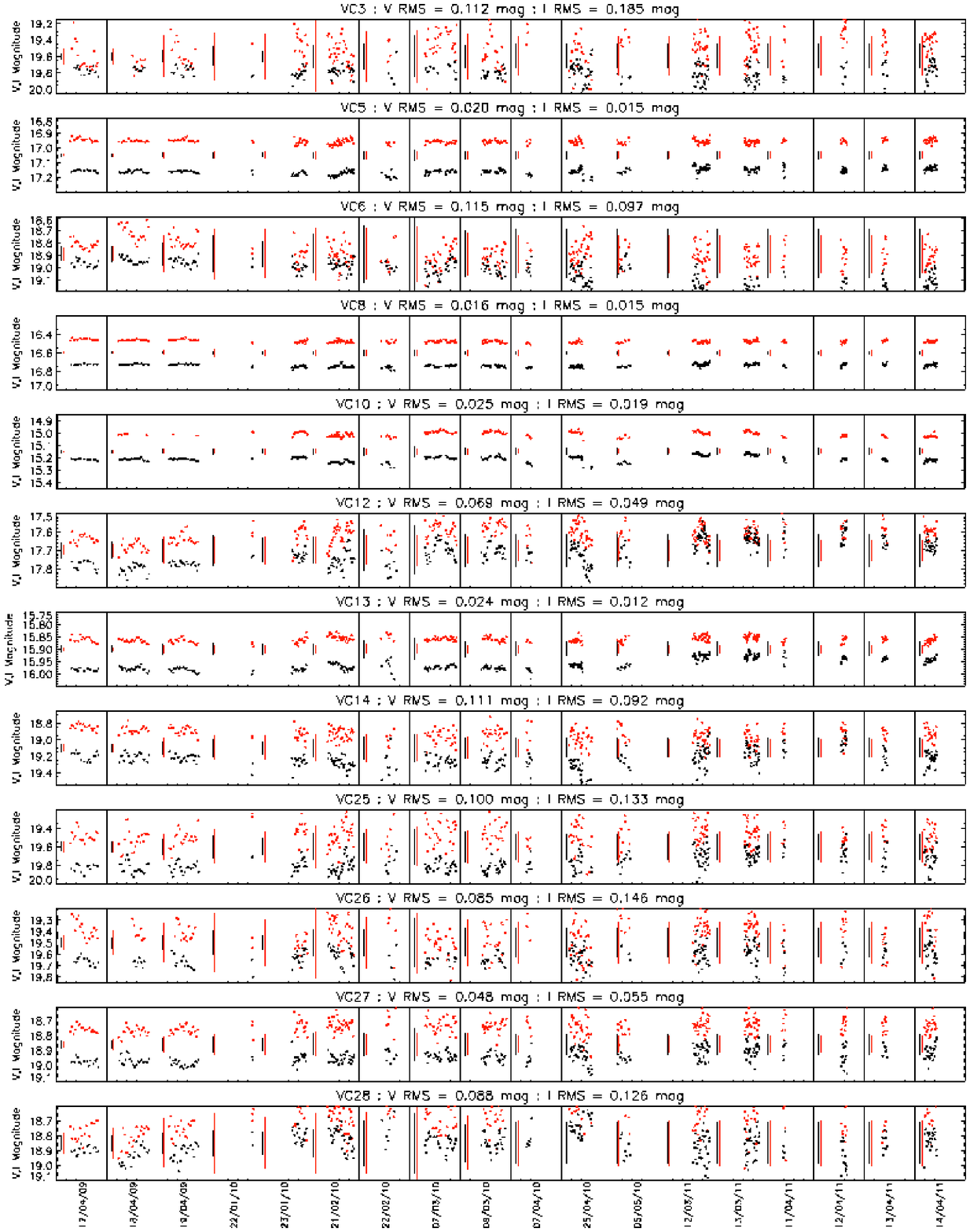,angle=0.0,width=\linewidth}
\caption{Light curves from A11 and A12 for selected candidate variables from SS11. $V$ and $I$ filter light curves are plotted in black and red, respectively,
         with mean photometric uncertainties per data point on each night plotted at the beginning of a night as vertical bars. The magnitude
         scale for each candidate has been matched to the scale used in SS11, and the $I$ filter light curves have been offset in magnitude by an arbitrary amount
         for clarity.
         \label{fig:nonvar2}}
\end{figure*}

SS11 present 27 candidate variables ``that show definite light variations'' but which they are unable to classify. Using the celestial
coordinates from SS11, we have found the corresponding $V$ and $I$ light curves in our data base, except for
candidate VC9 where the {\tt DanDIA} pipeline has failed to measure the reference flux of this star in both the $V$ and $I$ filter
reference images.

Candidates VC1, VC2, VC7, VC11, VC15, VC16, VC17, VC21, VC22, and VC24 lie within $\sim$3.7, 4.0, 3.7, 20.4, 20.2, 3.5, 13.3, 4.4, 4.7, and 4.3~pix,
respectively, of the known RR Lyrae variables V33, V42, V64, V46, V72, V53, V63, V45, V9, and V51, respectively. Inspection of the SS11
light curves for these candidates reveals that they mimic segments of RR Lyrae light curves, and we believe that the variability
detected by SS11 for these stars is due to the systematic influence on the light curves from the nearby RR Lyrae stars.

Candidates VC18, VC19, and VC20 all lie within $\sim$10~pix of the same highly saturated star that is also in the vicinity of the previously
discussed candidates RR11 and RR13 (see Section~\ref{sec:rrlyrae}). We believe that the poor quality of the difference images in such
close proximity to a highly saturated star is the cause of the variations observed in the SS11 light curves for these stars.

In Figure~\ref{fig:nonvar2}, we plot our light curves for VC3, VC5, VC6, VC8, VC10, VC12-14, and VC25-28 using the same scale on the
magnitude axis as that used in SS11, and in Figure~\ref{fig:nonvar1}, we plot our light curve for V80\_7, which was flagged as a suspected
variable by SS11 during their investigation into the identification of V80. As expected, all of our light curves
show systematic trends at some level (usually the few per cent level). However, for all of the candidates except VC27 and VC28, the amplitudes of the
variations in our light curves are considerably smaller than the amplitudes of the variations seen in the SS11 light curves. Furthermore, the
variations seen in our light curves do not generally correlate between wavebands on a reasonable fraction of nights. We also note that 
VC25 and VC26 are close to a saturated star. Hence, on the basis of our data, we do not confirm the variability of
the stars VC3, VC5, VC6, VC8, VC10, VC12-14, VC25, VC26, and V80\_7, and we believe that the variations seen by SS11
are most likely the signature of systematic errors introduced during the reduction process.

For candidate VC27, SS11 present a light curve with variations of amplitude $\sim$0.15~mag, and our light curve is not of sufficient precision
to definitively rule out such variations. However, the variations that are seen in our light curve for VC27 do not always correlate between wavebands
(e.g. see the night of 19/04/2009). For candidate VC28, our light curve on any single night does not show the $\sim$0.3~mag variations seen in the
SS11 light curve, and again the variations do not always correlate between wavebands (e.g. see the night of 17/04/2009). Hence, we also do not confirm
the variability of the stars VC27 and VC28.

\section{New Variables In NGC~5024}
\label{sec:newvar}

\begin{figure}
\centering
\epsfig{file=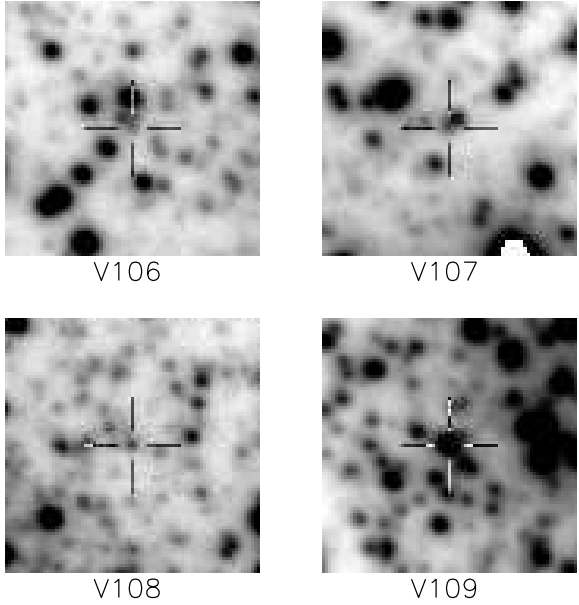,angle=0.0,width=\linewidth}
\caption{Finding charts constructed from our $V$ reference image; north is up and east is to the right. Each image stamp is of size
         23.7$\times$23.7 arcsec$^{2}$. Each confirmed variable lies at the centre of its corresponding image stamp and is marked by
         a cross-hair.
         \label{fig:find}}
\end{figure}

In Section~\ref{sec:sxphe}, we confirm the variable nature of the candidates SX4, SX16, and SX21 from SS11, and consequently, we assign them the new variable numbers
V106, V107, and V108, respectively. For future ease of identification, we provide the finding charts for these stars in Figure~\ref{fig:find}. We perform a frequency
analysis of the light curves of these new variables using {\tt PERIOD04} (\citealt{len2005}), the 
results of which we report in Table~\ref{tab:newvar}, along with their mean magnitudes, colours, and celestial coordinates.

\begin{figure}
\centering
\epsfig{file=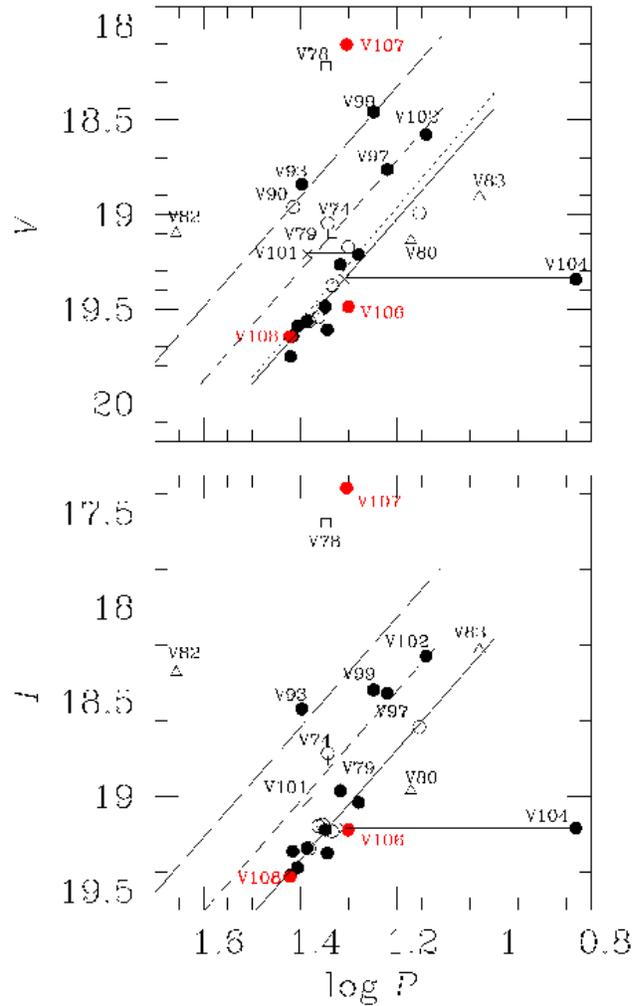,angle=0.0,width=\linewidth}
\caption{The period-luminosity relation in the $V$ and $I$ wavebands for the SX Phe stars in NGC~5024. The new SX Phe stars
         confirmed in this paper are plotted as solid red circles. The filled and empty black circles are the SX Phe stars discovered by
         A11 and \citet{jeo2003}, respectively. Empty squares and triangles are the SX Phe stars from \citet{dek2009}. The continuous
         straight line is the result of a least-squares fit to the stars pulsating in the fundamental mode. The short and long
         dashed lines correspond to the relations inferred for the first and second overtone pulsators, respectively, assuming
         the frequency ratios $F/1O = 0.783$ and $F/2O = 0.571$. The dotted line in the top panel is the P-L relation calculated by \citet{jeo2003}
         from six SX Phe in NGC~5024. See Figure~12 in A11 for more details.
         \label{fig:plrel}}
\end{figure}

We only detect one dominant frequency for each variable, which we use to plot the stars in Figure~\ref{fig:plrel} as solid red circles. This figure displays the P-L relation
for the SX Phe stars in NGC~5024 and it is an updated version of Figure~12 from A11. V106 and V108 follow the fundamental mode P-L relation for
the cluster and they also lie in the blue straggler region of the CMD (Figure 4 of A11). Hence we classify V106 and V108 as SX Phe stars
pulsating in the fundamental ($F$) mode. V107 lies above the P-L relations for the fundamental, first, and second overtone pulsators in the cluster,
and it also lies close to but outside of the blue straggler region in the CMD. However, V107 is blended with a brighter star (see Figure~\ref{fig:find})
and therefore it is likely that the reference flux for this variable is over-estimated. Assuming that V107 is actually an SX Phe variable, which
is an assumption consistent with its short period, then the blending could explain why it is brighter than the cluster P-L relation and why its colour places it outside of
the blue straggler region. We therefore tentatively classify V107 as an SX Phe variable for which we cannot speculate on the mode(s) in which it is pulsating.

\begin{table*}
\caption{Properties of the new variable stars in NGC~5024. We list the mean $V$ magnitude, mean $V-I$ colour, detected frequency $f$, $V$-waveband photometric 
         peak-to-peak amplitude $A$,
         and corresponding period $P = f^{-1}$ in columns 2, 3, 4, 5, and 6, respectively. The celestial coordinates at the epoch of our $V$ reference image
         (heliocentric Julian date $\sim$2455249.332~d) are listed in columns 7 and 8. The variable type is listed in column 9 along with the pulsational mode
         for the SX Phe variables in column 10. The numbers in parentheses indicate the uncertainty on the last decimal place.
         }
\centering
\begin{tabular}{cccccccccc}
\hline
Variable     & $V$    & $V - I$   & $f$         & $A$       & $P$          & RA          & Dec.       & Classification & Pulsation \\ 
Star ID      & (mag)  & (mag)     & (d$^{-1}$)  & (mag)     & (d)          & (J2000.0)   & (J2000.0)  &                & Mode(s)   \\
\hline
V106         & 19.487 & 0.310     & 20.01734(6) & 0.212(18) & 0.0499567(2) & 13 12 48.69 & 18 10 10.3 & SX Phe         & $F$ \\
V107         & 18.101 & 0.732     & 20.16156(8) & 0.050(6)  & 0.0495993(2) & 13 12 56.38 & 18 11 05.3 & SX Phe$^{a}$   & ? \\
V108         & 19.645 & 0.219     & 26.35550(8) & 0.121(12) & 0.0379427(2) & 13 12 59.52 & 18 11 17.4 & SX Phe         & $F$ \\
V109         & 15.208 & 1.117     & 0.04560(4)  & 0.050(5)  & 21.93(2)     & 13 12 53.55 & 18 09 55.4 & SR             & - \\
\hline
\end{tabular}
\raggedright
\\$^{a}$Uncertain classification.
\label{tab:newvar}
\end{table*}

In Sections~\ref{sec:eclbin}~\&~\ref{sec:sxphe}, we tentatively confirm the variable nature of the candidates W11 and SX12 from SS11, but since our detection of the
variability is border-line in these cases, we refrain from assigning these stars a V-number or speculating on their type of variability.
Further follow-up observations of better precision will be required to properly confirm the variable nature of these particular stars.

\begin{figure}
\centering
\epsfig{file=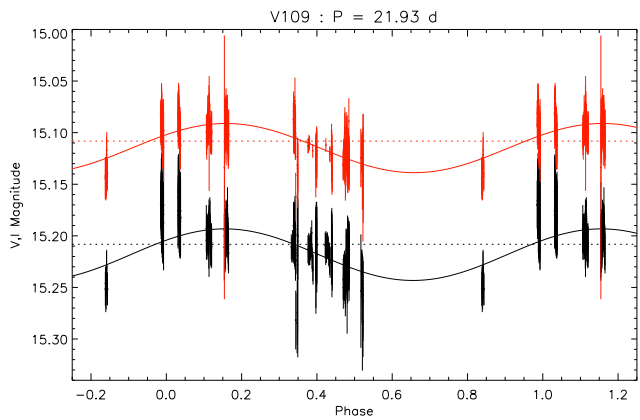,angle=0.0,width=\linewidth}
\caption{Light curve from A11 and A12 for V109 phase-folded using a period of 21.93~d. $V$ and $I$ filter observations are plotted in black and red, respectively.
         The $I$ filter light curve has been offset from the $V$ light curve by 0.1~mag for clarity. The fitted constant and sine-curve models are plotted
         as dashed and continuous curves, respectively.
         \label{fig:v109}}
\end{figure}

Finally in this Section, we report the discovery that the variable candidate VC10 from SS11 is actually a long period variable.
Although we have ruled out the short term variations of $\sim$0.3~mag amplitude claimed by SS11 (see their Figure~13),
a careful inspection of our corresponding light curve has revealed variations on the timescale of days
with an amplitude of a few percent (discernable between observation nights in Figure~\ref{fig:nonvar2}). We have performed a period search in the range 10-1000~d
again using {\tt PERIOD04} and we find the period with the strongest peak in the frequency spectrum is 21.93$\pm$0.02~d. However, it is quite possible
that the period we have found is an alias of the true period since our phase coverage at the timescale of days is quite low.

In Figure~\ref{fig:v109}, we plot the phase-folded light curve of VC10 at the period of 21.93~d using black and red points for the $V$ and $I$ photometric measurements, respectively.
For clarity, we offset the $I$ light curve from the $V$ light curve by 0.1~mag. Clearly the photometric observations are well correlated between the two wavebands.
Furthermore, a sine-curve fit at the derived period of 21.93~d yields similar amplitudes of $\sim$0.050 and 0.048~mag in the $V$ and $I$ wavebands, respectively,
and $\Delta \chi^{2}$ values of $\sim$150.6 and 169.7, respectively. Hence there is no doubt that this star is exhibiting low-amplitude long-period 
true variability. The star lies neatly on the red giant branch of the CMD, and we classify it as a semi-regular (SR) type variable. We assign the
variable number V109 and report the variable star properties in Table~\ref{tab:newvar}.

We comment that for V109, the variable nature of this star was not detected by SS11 for the simple reason that it is impossible to detect
low-amplitude smooth variations at a 21.93~d period with only $\sim$7.2~h of continuous observations. Therefore the discovery of V109
should be attributed to the work in this paper, as opposed to the discovery of V106-V108, which should be attributed to SS11.

In Table~\ref{tab:lc}, available in full in electronic form (see Supporting Information),
we provide our $V$ and $I$ time-series photometry for all the SX Phe and long-period variables in NGC~5024 from the works of A11, A12, and this
paper\footnote{The light curves for all the RR Lyrae stars in NGC~5024 have already been published in A11 and A12.}.

\begin{table*}
\caption{Time-series $V$ and $I$ photometry for all the SX Phe and long-period variables in NGC~5024 from the works of A11, A12, and this paper.
         The standard $M_{\mbox{\scriptsize std}}$ and instrumental $m_{\mbox{\scriptsize ins}}$ magnitudes are listed in columns 4 and 5,
         respectively, corresponding to the variable
         star, filter, and epoch of mid-exposure listed in columns 1-3, respectively. The uncertainty on $m_{\mbox{\scriptsize ins}}$ is listed
         in column 6, which also corresponds to the uncertainty on $M_{\mbox{\scriptsize std}}$. For completeness, we also list the quantities
         $f_{\mbox{\scriptsize ref}}$, $f_{\mbox{\scriptsize diff}}$, and $p$ from Equation~1 in A11 in columns 7, 9, and 11,
         along with the uncertainties $\sigma_{\mbox{\scriptsize ref}}$ and $\sigma_{\mbox{\scriptsize diff}}$ in columns 8 and 10.
         This is an extract from the full table, which is available with the electronic version of the article (see Supporting Information).
         }
\centering
\begin{tabular}{ccccccccccc}
\hline
Variable & Filter & HJD & $M_{\mbox{\scriptsize std}}$ & $m_{\mbox{\scriptsize ins}}$ & $\sigma_{m}$ & $f_{\mbox{\scriptsize ref}}$ & $\sigma_{\mbox{\scriptsize ref}}$ & $f_{\mbox{\scriptsize diff}}$ &
$\sigma_{\mbox{\scriptsize diff}}$ & $p$ \\
Star ID  &        & (d) & (mag)                        & (mag)                        & (mag)        & (ADU s$^{-1}$)               & (ADU s$^{-1}$)                    & (ADU s$^{-1}$)                &
(ADU s$^{-1}$)                     &     \\
\hline
V67 & $V$ & 2454940.17574    & 14.131 & 15.410 & 0.001  & 5984.296  & 1.689  & 1058.367 & 3.975  & 1.2186 \\
V67 & $V$ & 2454940.19142    & 14.128 & 15.408 & 0.001  & 5984.296  & 1.689  & 1081.793 & 3.922  & 1.2230 \\
\vdots    & \vdots & \vdots  & \vdots & \vdots & \vdots & \vdots    & \vdots & \vdots   & \vdots & \vdots \\
V67 & $I$ & 2455219.49351    & 12.922 & 14.312 & 0.001  & 18675.923 & 3.782  & 131.660  & 12.956 & 0.7850 \\
V67 & $I$ & 2455219.50345    & 12.919 & 14.310 & 0.003  & 18675.923 & 3.782  & 21.386   & 5.951  & 0.1031 \\
\vdots    & \vdots & \vdots  & \vdots & \vdots & \vdots & \vdots    & \vdots & \vdots   & \vdots & \vdots \\
\hline
\end{tabular}
\label{tab:lc}
\end{table*}

\section{Discussion}
\label{sec:dis}

SS11 consider a combination of several statistics to select their 74 candidate variable stars. For each of their $\sim$10$^{4}$ light curves, they 
calculate an alarm statistic $A$ (\citealt{tam2006}), an excess variance statistic $\sigma_{\mbox{\scriptsize XS}}$ (\citealt{vau2003}), the Lomb periodogram significance level
$F$ at the best-fit period (\citealt{lom1976}), and the RMS magnitude deviation $r$. They then devise the following combined criterion
for selecting variable star candidates:
\begin{equation}
\begin{cases}
A > 1.0 \\
\sigma_{\mbox{\scriptsize XS}} > 0.09 \\
F < 10^{-4} \\
r > 0.01 \\
\end{cases}
\label{eqn:thresh}
\end{equation}
which selects 310 candidate variables. SS11 then rejected variable star candidates within 20~pix of the image edges and
within 10~pix of known variable stars (although this last rejection criterion was obviously not applied properly), and used
visual inspection to make the final selection of 74 candidates.

As pointed out in A12, the alarm statistic $A$ is flawed for variable star detection because of the normalisation by $\chi^{2}$ in its definition. Large 
amplitude variable stars have long runs of large residuals, but also a very high $\chi^{2}$ statistic, meaning that true large-amplitude variables yield undesirably small
values for $A$. Therefore we are sceptical that the constraint on the $A$ statistic applied in SS11 was actually useful for discriminating variable
from non-variable stars.

Both the Lomb periodogram significance level $F$ and the excess variance statistic $\sigma_{\mbox{\scriptsize XS}}$ are highly dependent on the accuracy of the
estimated uncertainties on the photometric data points. Under-estimation of the uncertainties leads to smaller values for $F$ and larger values for 
$\sigma_{\mbox{\scriptsize XS}}$, and vice versa. Since there is no discussion in SS11 of how the
thresholds on $F$ and $\sigma_{\mbox{\scriptsize XS}}$ were chosen, it is not clear how appropriate the adopted thresholds are.

The RMS diagram presented in Figure~8 of SS11 shows that $r > 0.01$~mag for all stars fainter than $\sim$17.5~mag in the $R$-waveband. Therefore the threshold on $r$
chosen by SS11 does not discriminate variable stars from non-variable stars except for the brightest stars in the sample. When searching for variability via the $r$ 
statistic, it is more appropriate to set a magnitude depedendent threshold determined from the RMS diagram itself (see the Vidrih index defined in Section~3.1 of 
\citealt{bra2008b}) since the RMS magnitude deviation is a magnitude dependent quantity that is generally larger for fainter stars.

Setting the variability detection threshold is a very important process that can be used to minimise the rate of false positive detections and maximise the number of 
detections of true variables. Our investigation into the candidate variables presented by SS11 clearly shows that the vast majority of their candidates are false positive 
detections with only a few detections of true variability, which means that they set their detection threshold too low. In fact, in SS11, there is no assessment of the
impact of random noise, the accuracy of the estimated photometric uncertainties, and the systematic trends in the light curves on the actual detection thresholds that should be applied in
order to minimise the number of false positives. Assuming that the estimated photometric uncertainties are correct and that there are no systematic trends in the
light curves, then Monte Carlo simulations employing the photometric uncertainties on the light curve data points may be used to determine the detection threshold
required to achieve a desired rate of false positives. However, in practice, the estimated photometric uncertainties are usually inaccurate and systematic trends almost certainly exist
in the light curves (as evidenced already in the light curve plots from both SS11 and this paper). Therefore, an appropriate detection threshold can only be determined from
analysis of the distribution of the detection statistic(s) for the light curves of the ensemble of stars under the assumption that the majority of stars 
are non-variable.

\section{Conclusions}
\label{sec:conclusions}

We have investigated the 74 candidate variable stars presented by SS11. We could not investigate four of these candidates
because they are either too faint (three cases) or the {\tt DanDIA} software failed to measure a reference flux on either
reference image (one case). Of the remaining 70 candidates, we find that 14 are true variable stars, 10 of which
were already discovered in previous works (although only 2 of these were available in the literature at the time of the
SS11 paper submission), and four of which are new discoveries. We assign the
variable star numbers V106-V109 to the newly discovered variables. We classify V106-V108 as SX Phe type variables whose discovery
is attributed to SS11, and we classify V109 as an SR type variable discovered in this paper.
We also tentatively confirm the presence of true variability in two other candidates.

From the analysis of our data, we strongly suggest that the other 54 candidate variable stars (or the majority of such)
presented by SS11 are false positive detections
where systematic trends in the light curves have been mistaken for true variability. For 17 of these cases, we can explain the systematic
trends as having been caused by their proximity to known RR Lyrae variables which are much brighter, and for 8 of these cases, the trends are
due to their proximity to saturated stars which reduce the local quality of the difference images. For the remaining 29 candidates,
investigation of our light curves does not yield any evidence of true variability at the amplitudes reported by SS11, and any lower amplitude
variations that we do see do not generally show the correlations between wavebands typical of true variability, which
leads us to conclude that in these cases, the variations seen in the SS11 light curves are due to trends
introduced during the reduction process.

\section*{Acknowledgements}

AAF is thankful of the support from the DGAPA-UNAM grant through project IN104612.
We would like to thank Christine Clement for her helpful comments on the SS11 paper.

\section*{Appendix A}
\label{app:append_A}

Briefly in this Appendix, we clarify some mistakes in the literature regarding variable star identification and astrometric coordinates.

In Table~2 of A11, there is a typo in the coordinates for V43. The coordinates should read: RA~$ = 13^{\mbox{\scriptsize h}} 12^{\mbox{\scriptsize m}} 53.08^{\mbox{\scriptsize s}}$,
Dec.~$= +18\degr 10\arcmin 55.5\arcsec$, J2000.

In A11, we did not report the celestial coordinates of V53, which is highly blended. For completeness, we have measured these coordinates on the difference images where the
variable star is not blended and we report them here as: RA~$ = 13^{\mbox{\scriptsize h}} 12^{\mbox{\scriptsize m}} 55.85^{\mbox{\scriptsize s}}$,
Dec.~$= +18\degr 10\arcmin 35.5\arcsec$, J2000. These coordinates correspond to the epoch of our $V$ reference image which is the heliocentric Julian date 
$\sim$2455249.332~d.

We note that SS11 arrive at the same conclusion as A11 about the previous misidentifications in the literature of the variables V57 and V72, and the non-variability of V81, V82, and V83.

\section*{Supporting Information}

Additional Supporting Information may be found in the online version of this article.

{\bf Table 2.} Time-series $V$ and $I$ photometry for all the SX Phe and long-period variables in NGC~5024 from the works of A11, A12, and this paper.

Please note: Wiley-Blackwell are not responsible for the content or functionality of any supporting materials supplied by the authors. Any queries
(other than missing material) should be directed to the corresponding author for the article.

\label{lastpage}

\end{document}